\documentclass{aip-cp}

\usepackage[numbers,sort&compress]{natbib}
\usepackage{rotating}
\usepackage{graphicx}
\usepackage{savesym}
\savesymbol{iint}
\savesymbol{iiint}
\savesymbol{iiiint}
\savesymbol{idotsint}
\usepackage{amsmath}
\restoresymbol{AMS}{iint}
\restoresymbol{AMS}{iiint}
\restoresymbol{AMS}{iiiint}
\restoresymbol{AMS}{idotsint}
\usepackage{dsfont}
\usepackage{url}
\usepackage{tablefootnote}
\begin{document}

\title{Random Matrix Ensembles For Many-Body Quantum Systems}

\author[aff1]{Manan Vyas}

\author[aff1]{Thomas H. Seligman}

\affil[aff1]{Instituto de Ciencias F{\'i}sicas, Universidad Nacional Aut{\'o}noma de M{\'e}xico, 62210 Cuernavaca, M{\'e}xico}

\maketitle

\begin{abstract}

Classical random matrix ensembles were originally introduced in physics to approximate quantum many-particle nuclear interactions. However, there exists a plethora of quantum systems whose dynamics is explained in terms of few-particle (predominantly two-particle) interactions. The random matrix models incorporating the few-particle nature of interactions are known as embedded random matrix ensembles. In the present paper, we provide a brief overview of these two ensembles and illustrate how the embedded ensembles can be successfully used to study decoherence of a qubit interacting with an environment, both for fermionic and bosonic embedded ensembles. Numerical calculations show the dependence of decoherence on the nature of the environment. 

\end{abstract}

\section{Introduction}

Classical random matrix ensembles are ensembles that describe the quantum representation of classical chaotic systems (quantum chaos) \cite{Ca-80, Be-81, Bo-84, Be-85, Si-01, Si-02, Br-04, Br-05}, but if considered for many-body systems they imply many-body interactions. However, Hamiltonians of quantum many-body systems should incorporate the few-body nature of interactions among its constituents. Thus, it is more appropriate to model these systems by embedded ensembles for fermions and bosons. In the present paper, we discuss how to make the random matrix ensembles for Hamiltonians of few- and many-body systems more realistic taking into account the two- or few-body character of interactions. The central point is the fact that many-body systems are mainly governed by two-body interactions, while the classical Gaussian ensembles correspond to $m$-body interactions, where $m$ is the total number of particles. A reasonable random matrix model might assume a Gaussian ensemble for the two-body interaction \cite{Fr-70, Bo-71}, a one-body part reflecting a mean field or the field of the nuclei in atoms or molecules and possibly some three-body interaction \cite{Ze-09}. In Random Matrix Theory (RMT) such ensembles are known as embedded ensembles. They account for few-body nature of interactions.

The principle of ergodicity allows one to compare spectral averages from a given spectrum with ensemble averages obtained using random ensembles. The early work dealt mainly with fermions \cite{Fr-70, Bo-71, Br-81} with findings that confirmed that the classical ensembles (consisting of $m$-body interactions) was a fair approximation for the fluctuations if spectra were unfolded individually \cite{Br-81}. In a more recent update of this work \cite{Fl-01}, it was shown that actually the first and second moments of the spectra show strong non-ergodicity or at least very slow convergence. This fact does not surprise because already French \cite{Fr-73} showed that the convergence for the spectral density goes with $1/\log(N)$, where $N$ is the dimension of the resulting Hamiltonian matrix. For two-level bosonic systems, the situation is more involved \cite{Be-01}. 

In addition to two-body interactions, realistic Hamiltonians are also characterized by a variety of symmetries (in addition to the number of particles). For instance, spin is a good symmetry for mesoscopic systems, parity is a good symmetry for nuclei, and so on. There have been a variety of studies exploring varied aspects of embedded ensembles with and without symmetries. For further details, we refer the readers to the following papers: bosonic embedded ensembles \cite{Be-01, Be-01a, As-01, As-02, Pa-02, Be-03, Ku-03, Be-03a, Be-11, Ma-12a, Ma-12b, De-13, Ma-15, Ma-15a, Ma-16} and fermionic embedded ensembles \cite{Mo-75, Fr-88, Fl-94, Fl-96, Fl-97, Ho-95, Ze-96, Ja-97, Ge-97, Al-00a, Ma-08, Pa-08, Ma-09, Ma-10a, Ma-10b, Ma-11a, Ma-11, Ma-12, Ma-12b, Ma-15, Ma-15a, Ma-16, Be-16}. We restrict our discussion in the present paper to embedded ensembles without symmetries. We will only marginally discuss embedded ensembles of distinguishable particles in the context of quantum chaos and its meaning for many-body systems.

Non-equilibrium dynamics of quantum systems has not been understood in much detail. Few attempts have been made to understand thermalization \cite{Ma-11}, transport properties \cite{Ma-15, Be-16} and behavior of survival probability and entropy \cite{Ma-17,Ma-16} using embedded fermionic and bosonic ensembles. Going beyond these, we consider the decoherence of a qubit coupled to a random many-body environment. The composite state is described by density matrix. As qubit is interacting with environment, the reduced density matrix describes the state of qubit. The random environment is modeled by realistic fermionic and bosonic embedded ensembles and results are compared with random GOE environment.

The term "quantum chaos" frequently employed in the context of the use of RMT in quantum mechanical problems  was often interpreted as “quantum manifestations of classical chaos” and considerable progress has been achieved in this direction using periodic orbits and semi-classical methods. Great progress has been made along these lines in general terms and for quantum graphs \cite{Do-92, Ko-99}. Pluhar and Weidenm{\"u}ller have recently made great progress showing that a complete quantum graph with pairwise incommensurable edges in the large edge-number limit has the same joint probability density as the corresponding random matrix ensemble \cite{Pl-15}. This gives the first ensemble of systems that has the properties of quantum chaos. Nevertheless we have to note that it is a one-body system. Yet Wigner \cite{Wi-55} introduced RMT into physics precisely to describe complicated many-body physics, and many-body physicists have rather systematically applied the term in Wigner's sense, to any system that experimentally or numerically displays RMT spectral statistics, whether it has a classical limit or not. In the context of fermionic and bosonic ensembles such limits are indeed subject of discussion, which we will not go into. In this context, systems of distinguishable particles or spins and the corresponding embedded ensembles \cite{Piz-08} have been used \cite{Go-16}. We shall not enter this little developed field, but these systems have gained increasing relevance in applications of 1-D systems and with the recent analytic demonstration of Prosen \cite{arXiv17} that for a particular, but rather generic, periodically time dependent system with one and two-body interactions of distinguishable spins we obtain an RMT two-point function. We shall return to the matter of quantum chaos in the next section. 

This is the third ELAF set of lectures given. It is worth while to mention that the tradition is old and venerable as M. L. Mehta gave a series in the nineteen seventies. The two more recent ones discuss fidelity studies \cite{Pi-08} and wishart ensembles of singular matrices \cite{Vi-14} and may well be used as complementary material. In the present lectures, to be self-contained, we give a brief summary of the basics of RMT and quantum chaos in the next sections, but for many applications we rely on rather extensive references given both above and throughout the paper, in the firm belief that an extended set of references is very helpful to the interested students. We define, in the fourth section, embedded ensembles for many-body fermionic and bosonic systems incorporating the few-body nature of the interactions. We compare and contrast the known results for one-point functions (eigenvalue density) and two-point functions (fluctuation measures) for these ensembles. In the fifth section we present new results for decoherence in a single qubit system in the presence of an environment described by fermionic and bosonic embedded ensembles. Finally, we give conclusions and an outlook.

\section{Classical Random Matrix Ensembles}

RMT has been successfully used in diverse areas, such as econophysics, nuclear physics, quantum chaos, wireless communications, number theory, quantum information science, quantum chromodynamics and so on, with wide ranging applicability to various mathematical, physical and engineering branches \cite{Br-81, Gu-98, Pa-07, Wr-10, Ba-10, Fo-10, Ak-11, Co-12, Ko-14, Bo-16}. Random matrix ensembles were first introduced by Wishart in multivariate analysis  \cite{Wi-28}. Eli Cartan \cite{Ca-35} gave a systematic description of the "classical" ensembles, while the aspects of multivariate analysis were further developed by Pastur \cite{Pa-67}. However, their extensive application began with pioneering work of Wigner in 1955 to explain neutron resonance data \cite{Wi-55}. Mathematical foundations were laid by Hua \cite{Hua-63}, Mehta and Dyson \cite{Dy-62a, Dy-62b, Dy-62c, Me-04}. RMT helps to analyze statistial properties of physical systems whose exact Hamiltonian is too complex to be studied directly. The exact Hamiltonian of the system in consideration is represented by an ensemble of random matrices that incorporate the global symmetry properties (namely spin and time-reversal invariance) of the system. If $T$ denotes the time-reversal operator, it turns out that it is an involutional anti-unitary operator with $T^2=\pm 1$ \cite{Sa-87}. Depending on number of half-integer spins and time-reversal ($T$) invariance properties, the classical random matrix ensembles are classified into three classes: Gaussian Orthogonal Ensembles (GOE), Gaussian Unitary Ensembles (GUE) and Gaussian Symplectic Ensembles (GSE).

\begin{table}[h!]
  \centering
  \caption{Classical random matrix ensembles. Here, $t$ denotes `transpose' and $*$ denotes `complex conjugation' and $\dagger$ denotes `conjugate transpose' of the matrix respectively. For GSE, $(\sigma_1, \sigma_2, \sigma_3)$ are Pauli spin matrices and ${\textrm I}$ is the identity matrix.}
\scalebox{0.78}{
  \begin{tabular}{|l|c|c|c|}
    \hline
    & GOE & GUE & GSE \\
    \hline
    Symmetries& integer spins and/or even number of half-integer spins & any spin & odd number of half-integer spins \\
                        & $T$ is preserved ($T^2=1$) & $T$ is not preserved & $T$ is preserved ($T^2=-1$) \\
    \hline
   Hamiltonian Structure & Real symmetric & Complex hermitian & Quaternion real \\
 & $H=H^t=H^*$ & $H=H^\dagger$ & $H=H_0 \; \textrm{I} + i \;\displaystyle\sum_{j=1}^3 \; H_j\; {\sigma_j}$ \\
 & &  & $H_0=(H_0)^t=(H_0)^*$\\
 & &  & $H_j=-(H_j)^t=(H_j)^*$\\
   \hline
 $N=2$ Example & $H = \left[ \begin{array}{cc} a+b & c  \\ c & a-b \end{array}\right]$ & $H = \left[ \begin{array}{cc} a+b & c-id  \\ c-id & a+b \end{array}\right]$ & $H_0=\left[ \begin{array}{cc} a & b  \\ b & c \end{array}\right]$, $H_1=\left[ \begin{array}{cc} 0 & -d  \\ d & 0 \end{array}\right]$ \\
 & & & $H_2=\left[ \begin{array}{cc} 0 & e  \\ -e & 0 \end{array}\right]$, $H_3=\left[ \begin{array}{cc} 0 & -f  \\ f & 0 \end{array}\right]$\\
 & $a,b,c \xrightarrow[]{\text{i.i.d.}} G(0,v^2)$ & $a,b,c,d \xrightarrow[]{\text{i.i.d.}} G(0,v^2)$ & $a,b,c,d,e,f \xrightarrow[]{\text{i.i.d.}} G(0,v^2)$\\
    \hline
  \end{tabular}}
  \label{tab:table1}
\end{table}

In finite dimensional Hilbert spaces, the Hamiltonian of a system can be represented by an $N \times N$ matrix; see Table \ref{tab:table1} for details. The matrix elements are chosen to be independent identically distributed (i.i.d.) Gaussian random variables with zero mean and variance $v^2$, i.e. $G(0,v^2)$. As the name suggests, these ensembles will be invariant under orthogonal [O($N$)], unitary [U($N$)] and unitary symplectic [USp($2N$)] transformations. It can be shown that the definition of these ensembles imply that the $H$'s should be real symmetric/complex hermitian/real quaternion for GOE/GUE/GSE in any representation differing from any other by an invariance under orthogonal/unitary/unitary symplectic transformations. Note that the matrix dimension must be even for GSE and all energy eigenvalues will be doubly degenerate (Kramer's degeneracy).  As an aside, GSE may prove to be appropriate for analyzing generic properties of spin lattice models for fermions with odd number of lattice sites. 

As the matrix elements are i.i.d. $G(0,v^2)$ for the classical ensembles, it is possible to derive the normalized joint probability distribution function (jpdf) of all eigenvalues $E_i$, $i=1,\ldots,N$. The jpdf of eigenvalues is defined as \cite{Me-04},
\begin{equation}
\rho_{\beta}(E) = {\cal N}_\beta \displaystyle\prod_{i < j=1}^N |E_i - E_j|^\beta \; \exp\left(  - \displaystyle\frac{\beta}{4} \displaystyle\sum_{i=1}^N E_i^2\right) \;.
\label{eq-1a}
\end{equation}
Here, ${\cal N}_\beta$ is the normalization constant and the $(N \times N)$ $H$ matrix is real symmetric, complex hermitian or quaternion real for Dyson parameter $\beta=1$, 2, 4. In order to derive Eq. \ref{eq-1a}, we first need to integrate over all i.i.d. matrix elements $H_{ij}$ (real, complex or quaternion real) in the equation for jpdf for matrix elements, $\int dH \exp(-\beta \mathrm{Tr}(HH^\dagger)/4v^2)$ where $^\dagger$ represents transpose conjugation. For $\beta=1$, $\mathrm{Tr} H^2 = \sum_i H_{ii}^2 + 2 \sum_{i<j} H_{ij}^2$. Thus, ${\overline{H_{ij}^2}} = (1+\delta_{ij}) v^2$ and the mean is zero. Here the bar denotes ensemble average. As matrix elements are chosen to be i.i.d. $H_{ii} \in G(0,2v^2)$ and $H_{ij} \in G(0,v^2)$ with $i \neq j$, the Gaussian weights in the jpdf for matrix elements factorise completely. The Vandermonde determinant $\prod_{i < j=1}^N |E_i - E_j|^\beta$ in the jpdf for eigenvalues (Equation \eqref{eq-1a}) comes from the Jacobian of transformation from the matrix elements space to the eigenvalue space. Without loss of generality, we choose $v^2=1$.

One can show that the limiting distribution of eigenvalue density of these classical Gaussian random matrix ensembles follows Wigner's semi-circle law \cite{Wi-55}. Importantly, the semi-circle eigenvalue density for classical random ensembles is independent of the nature of the distribution of its matrix elements if it has finite moments. Assume a suitable distribution $D$ of matrix elements of a family of symmetric $N$-dimensional random matrices. If $D$ has finite moments of all orders, then the distribution of eigenvalues ($E_i$, $i=1,\ldots,N$) averaged over all the random matrices converges to a semicircle distribution, $\rho(E) \xrightarrow{N \to \infty} 2\sqrt{R^2-E^2}/\pi R^2$ with $-R \leq E \leq R$ \cite{Wi-55}. The $p$-th order moment $M(E^{p}) = \int E^p \rho(E) dE$ of the semicircle distribution is given in terms of Catalan numbers $C_p$, $M(E^{2p}) = \left( \frac{R}{2} \right)^{2p} C_p$; $C_p=\frac{1}{p+1}\binom{2p}{p}$. Also, the range $R$ of the semicircle depends on the first two moments of  $D$. If we change the distribution from Gaussian to either exponential or uniform, we still obtain semicircle law for the eigenvalue distribution numerically, except for one outlier for both exponential and uniform distributions. See Figure \ref{dos} ahead for eigenvalue density of a 100 member GOE with $N=252$. Agreement of the numerical histogram with the semicircle is excellent. 

\section{Quantum Chaos}

As mentioned in the introduction, the term 'quantum chaos' is often applied in a more general and looser sense, namely, for complicated many-body problems where RMT statistics are found to hold quite well. A paradigmatic example has been the kicked Ising chain introduced by T. Prosen \cite{Pr-02} mainly for convenience as the angle between the kicked homogeneous magnetic field and the Ising interaction determines variation from integrability to situations where the spectral fluctuations are well represented by those of classical ensembles. Note that in such a system, we have to replace the energy spectrum by the quasi-energy spectrum, i.e. the spectrum of the unitary time evolution operator. It has a homogeneous spectral density (except for xenon-like very short times) and thus unfolding becomes irrelevant up to a trivial normalization. While general semi-classical theory may apply, here we have, for the first time, a specific proof of RMT behavior in a many-body system \cite{arXiv17}. The fact that the examples given for a system with distinguishable particles seems of minor importance. Rather it shows that the general ideas carry through, though exceptions will always occur. It also gives a firm basis to the use of embedded ensembles and might show a way around the no-ergodicity problems associated with the embedded ensembles for further analytic proofs.

RMT has been established to be one of the central themes in quantum physics with the recognition that quantum systems, whose classical analogues are chaotic, follow RMT. Early comments, e.g. by Percival and others, probably instigated by the harmonic oscillator, indicated that one should expect regular spectra for integrable systems and irregular ones for chaotic systems, Berry and Tabor \cite{Be-77} recognized the singularity of the oscillator case and stipulated that in a more generic context, one would expect integrable systems to have a random or Poisson behavior. It became increasingly clear that the irregularity included level repulsion \cite{Mc-79}. Berry \cite{Be-81}, though the study lacked the numerical exactitude to determine the quantum behavior of a Sinai biliard, proposed in some meetings the {\it Quantum Chaos} (QC) conjecture that states that we expect, in the semi-classical limit, spectral fluctuations of the classical random matrix ensembles for quantized ergodic Kolmogorov systems (the systems commonly called chaotic if we consider the subspectra belonging to each irreducible representation on any symmetry group of the system separately). The ensemble will essentially be determined by symmetry considerations. In \cite{Ca-80}, this conjeture was introduced explicitly and discussed for the GOE case, while in \cite{Bo-84, Bo-84a} the QC conjecture was formulated in detail and numerical evidence of improved quality for the Sinai Billard was obtained. Numerical evidence for the GUE case was given in \cite{Se-85} for homogeneous quartic Hamiltonians with a magnetic field. Rather extensive evidence for this conjecture has been assembled since and semi-classical considerations by Berry \cite{Be-85, Be-89} and others \cite{Br-04, Br-05, Ke-07, He-07} has been given. Yet exceptions exist \cite{Bo-96} and the conditions that would have to be introduced for a proof have been elusive \cite{Le-92, Si-93, Le-91}. Nevertheless this conjeture has been applied successfully and it is widely accepted.

\begin{figure}[ht]
  \centerline{\includegraphics{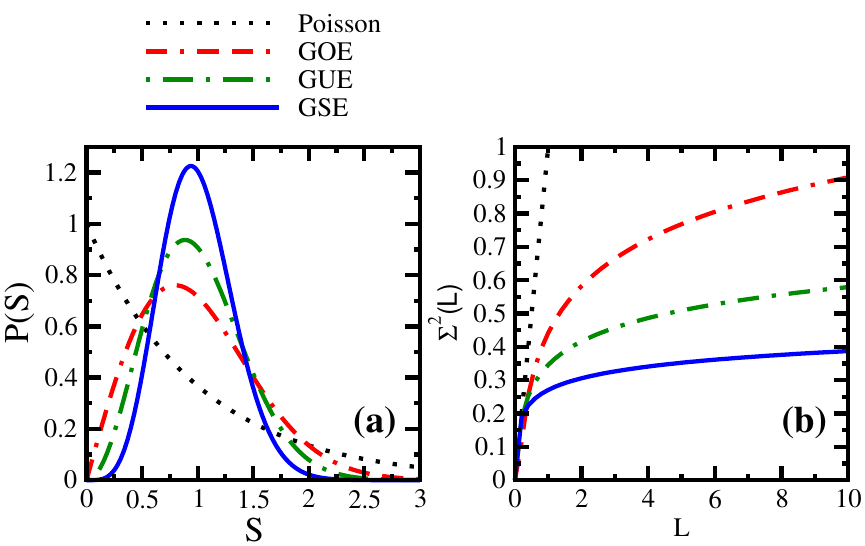}}
  \caption{(a) NNSD and (b) $\Sigma^2(L)$, for GOE, GUE, GSE compared with Poisson distribution.}
\label{nnsd}
\end{figure}

The spectral fluctuation properties are usually characterized by the nearest neighbor spacing distribution (NNSD) \cite{Ne-29} and by the Dyson-Mehta $\Delta_3$ statistic or the number variance $\Sigma^2$ \cite{Dy-63}. As the average spectral density of any given physical system is not given by RMT, one needs to rescale the spacing between the eigenvalues according to the local average eigenvalue density. This is known as 'unfolding' of the eigenvalue spectrum. For a sequence of eigenvalues $E_i$, $i=1,\ldots,N$ and average spectral density ${\overline{\rho(E)}}$, the unfolded sequence is ${\overline E}_i = \int_{-\infty}^{E_i} {\overline{\rho(E)}} dE$. The resulting unfolded spectrum has unit average level spacing. The NNSD $P(S)$ gives the statistics of spacings between adjacent unfolded eigenvalues. It is a histogram of neighboring unfolded levels. The number variance $\Sigma^2(r)$ is defined as the variance of the number of unfolded eigenvalues in an interval of length $r$. It is obtained by counting the number of levels in a sequence that contains $r$ levels on average in an energy interval when moving along the spectrum. It is an exact two-point measure i.e. it contains the same information as the two-point function of the unfolded spectrum. For a GOE, it is given by $\Sigma^2(r) = \frac{2}{\pi^2} (\log 2\pi r + 1 + \gamma - \frac{\pi^2}{8})$; $\gamma$ is the Euler-Mascheroni constant. Note that this is an exact result valid at the center of the semi-circle. The $\Delta_3(L)$ statistic is the mean-square deviation of the distribution function ($\int_{-\infty}^{E} \rho_\beta(x) dx$) of the unfolded spectrum from the best fit straight line; $L$ is the length of the energy interval over which $\Delta_3(L)$ is calculated. In fact, the $\Delta_3$ statistic is an integral of the number variance $\Sigma^2(r)$, $\Delta_3(L) = \frac{2}{L^4} \int_0^L (L^3-2L^2r+r^3) \Sigma^2(r) dr$. Therefore, $\Delta_3(L)$ is much smoother than $\Sigma^2(r)$ and hence, widely used in the literature. Yet, it is not an exact two-point measure \cite{Ve-87}. For GOE, it is given by $\Delta_3(L) = \frac{1}{\pi^2} (\log (2\pi L) + \gamma - \frac{5}{4} - \frac{\pi^2}{8})$. For example, the NNSD showing von-Neumann Wigner ‘level repulsion’ [$P(S) \propto S^\beta \exp(-a_\beta S^2)$] (also known as {\it Wigner surmise}) and $\Delta_3$ statistic showing ‘spectral rigidity’ [$\Delta_3(L) \propto \log(L)$] are exhibited by quantum systems well described by RMT. The number variance $\Sigma^2(r)$ displays similar behavior [$\Sigma^2(r) \sim 2 \log(r)/\beta\pi^2$]. For integrable systems, NNSD follows Poisson behavior [$P(S) = \exp(-S)$] and the number variance as well as $\Delta_3$ statistic exhibit linear behavior [$\Sigma^2(r) = r$, $\Delta_3(L) = L$] \cite{Be-77}. Note though that this is not entirely true for quantized integrable systems. At larger distances in the spectrum, the two measures will saturate, both for integrable \cite{Th-85} and chaotic \cite{Th-84, Th-85a} systems, though the NNSD is practically not affected. Berry explains this semi-classically by the existence of a shortest orbit in any given system. All these properties have been shown in 2D single particle systems, but the general argument from semi-classical approximations is expected to be valid whenever the semi-classical limit is meaningful and saturation becomes negligible (i.e. it occurs at practically irrelevant large spectral distances as particle number increases). These predictions have been confirmed in a huge amount of experimental, numerical and theoretical studies by now. For example, NNSD for an integrable semi-circular billiard follows Poisson distribution \cite{Bo-84}, NNSD and $\Delta_3$ for the chaotic Sinai billiard follows GOE \cite{Bo-84a}. Importantly, the Nuclear Data Ensemble \cite{Ha-82} follows RMT establishing that the neutron resonance region can be viewed as a region of chaos. As an example, we show the spacing distributions for integrable systems (Poisson) and chaotic sytems (GOE/GUE/GSE) in Figure \ref{nnsd}(a) and similarly for the number variance in Figure \ref{nnsd}(b) as a function of energy interval $L$. Now, we move on to define embedded ensembles and analyze their spectral properties.

\section{Embedded Gaussian Orthogonal Ensembles (EGOE)}

From now on we mainly discuss the orthogonal case but the unitary and symplectic cases are very similar. Note that there are no detailed studies on EGSE so far.

Constituents of finite many-body quantum systems such as nuclei, atoms, molecules, small metallic grains, quantum dots, arrays of ultracold atoms, and so on, interact via few-body (predominantly two-body) interactions. As is well-known, the classical random matrix ensembles (GOE) incorporate many-body interactions. Thus, it is more appropriate to use random matrix ensembles incorporating the few-body nature of interactions. The matrix ensembles generated by random few-body interactions are known as embedded Gaussian orthogonal ensembles [EGOE($k$)]  \cite{Fr-70,Bo-71}. These ensembles are generated by representing the few particle ($k$) Hamiltonian by a classical GOE and then the many-particle Hamiltonian ($m>k$) is generated by the Hilbert space geometry. In other words, $k$-particle Hamiltonian is emedded in the $m$-particle Hamiltonian as non-zero many-particle matrix elements are appropriate linear combinations of $k$-particle matrix elements. Due to few-body selection rules, the many-particle Hamiltonian has many zero matrix elements, unlike a GOE. Initially, when these EGOEs were introduced, all the work was done for fermionic systems. Later, notation 'BEGOE' was introduced for EGOEs for bosonic systems. We take the opportunity to simplify the notations and heceforth respresent fermionic EGOEs by FEGOE and bosonic EGOEs by BEGOE, here `F' and `B' stand for fermions and bosons respectively. This also provides advantage of introducing the notation `DEGOE' which represents EGOEs for distinguishable particles (particles which have fixed locations in space like nuclear skeletons of molecules often idealized as spin chains or spin networks). Note that the spin chains or spin networks do not embed the Hamiltonian in the defining $k$-particle spaces in the $m$-particle Hamiltonian, instead the few-body character of interactions is often accounted in the $m$-particle non-random Hamiltonian, such as tight-binding Hamiltonians. Some DEGOEs can also be visualized in terms of embedded ensembles that preserve spin projection quantum number $M_{\cal{S}}$. These may be useful in deriving generic results for entanglement measures.

Consider a system of $m$ identical spinless fermions[bosons] distributed in $\ell$ degenerate single particle (sp) levels with $k$-body interactions ($k \leq m$). The embedding algebra for FEGOE($k$)[BEGOE($k$)] is $SU(\ell)$. These ensembles are defined by three parameters $(\ell,m,k)$ and the random $k$-body Hamiltonian in second quantized form is,
\begin{equation}
H(k) = \displaystyle\sum_{\alpha,\;\gamma} \; v^{\alpha,\;\gamma}_k \; \alpha^\dagger(k) \; \gamma(k) \;.
\label{eq-1}
\end{equation} 
Here, $\alpha^\dagger(k)$ and $\gamma(k)$ respectively are $k$-particle creation and annihilation operators for fermions[bosons], i.e. $\alpha^\dagger(k) = \prod_{i=1}^{k} a^\dagger_{j_i}$[$\prod_{i=1}^{k} {\cal N}_{\alpha} \; b^\dagger_{j_i}$. Here, ${\cal N}_{\alpha}$ is the factor that guarantees unit normalization of $k$-particle bosonic states]. They obey the usual anti-commutation[commutation] relations. In Eq. \eqref{eq-1}, the sum stands for summing over a subset of $k$-particle creation and annihilation operators. We order the sp levels (denoted by $\mu_i$) in increasing order, $\mu_1 \leq \mu_2 \leq \cdots \leq \mu_k \leq \cdots \leq \mu_m$ in occupation number basis. In Equation \eqref{eq-1}, $v^{\alpha,\;\gamma}_k$ are anti-symmetrized[symmetrized] few-body matrix elements chosen to be randomly distributed independent Gaussian variables with zero mean and variance
\begin{equation}
{\overline{v^{\alpha,\;\gamma}_k \; v^{\alpha^\prime,\;\gamma^\prime}_k}} = v^2 \; \left( {\delta_{\alpha,\;\gamma^\prime}} {\delta_{\alpha^\prime,\;\gamma}} + {\delta_{\alpha,\;\alpha^\prime}} {\delta_{\gamma^\prime,\;\gamma}} \right) \;.
\label{eq-2}
\end{equation} 
We set $v=1$ without loss of generality. In other words, $v^{\alpha,\;\gamma}_k$ is chosen to be a $\binom{\ell}{k}$$\left[\binom{\ell+k-1}{k}\right]$ dimensional GOE in $k$-particle spaces. For example, the FEGOE(2) is defined by the two-body Hamiltonian $H(2) = \sum_{i<j,k<l} v_2^{k,l,i,j} a_l^\dagger a_k^\dagger a_i a_j$, with $v_2^{k,l,i,j} = \langle \mu_k \mu_l \mid H \mid \mu_i \mu_j \rangle$ being the two-particle anti-symmetrized matrix elements. Similarly, BEGOE(2) is defined by $H(2) = {\cal N} \sum_{i\leq j,k\leq l} v_2^{k,l,i,j} b_k^\dagger b_l^\dagger b_i b_j$ with $v_2^{k,l,i,j} = \langle \mu_k \mu_l \mid H \mid \mu_i \mu_j \rangle$ being the two-particle symmetrized matrix elements.

Each possible distribution of $m$ fermions[bosons] in the $\ell$ sp levels (with $\ell > m$ for fermions) generates a configuration or a basis state. Distributing the $m$ fermions[bosons] in all possible ways in $\ell$ levels generates the $d(\ell,m)=\binom{\ell}{m}$[$\binom{\ell+m-1}{m}$] dimensional Hilbert space or basis space. For fermions, this is similar to distributing $m$ particles in $\ell$ boxes with the conditions that the occupancy of each box can be either zero or one and the total number of occupied boxes equals $m$. For bosons, this is similar to distributing $m$ particles in $\ell$ boxes with the conditions that occupancy of each box lies between zero and $m$ and the maximum number of occupied boxes equals $m$. Given the sp states $| \mu_i\rangle$, $i=1,\;2,\ldots,\;\ell$ in the occupation number basis, the action of the Hamiltonian operator $H(k)$ defined by Equation \eqref{eq-1} on the many-particle basis states $| \mu_1 \mu_2 \ldots \mu_k \ldots \mu_m \rangle$ generates the FEGOE($k$)[BEGOE($k$)] ensemble in $m$-particle spaces.  For fermions, $\mu_i = 0$ or $1$ following Pauli's exclusion principle whereas $\mu_i$ can take any value between $0$ and $m$ for bosonic systems. Unlike GOE, FEGOE($k < m$) and BEGOE($k < m$) incorporate few-body nature of interactions that results in correlations between matrix elements of $H(k)$ and many of them will be zero due to $k$-body selection rules. By construction, the case $k = m$ is identical to a canonical GOE. 

\begin{figure}[ht]
  \centerline{\includegraphics[width=6.5in]{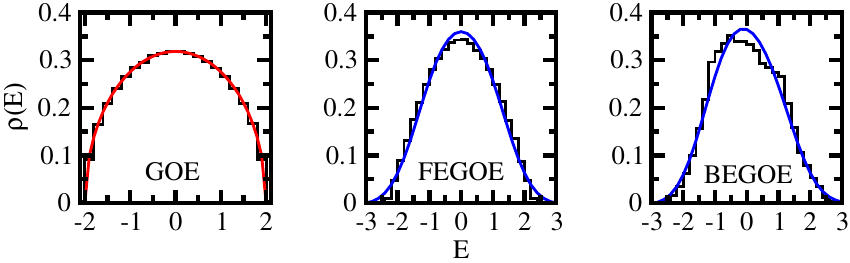}}
  \caption{Ensemble averaged eigenvalue density $\rho(E)$ for a 100 member GOE, FEGOE ($\ell=10$, $m=5$, $k=2$) and BEGOE ($\ell=2$, $m=251$, $k=2$). The numerical histograms are normalized to unity in all the plots. For GOE, comparison is made with a semicircle (solid curve) whereas for FEGOE and BEGOE, comparisons are made with ED corrected Gaussians (solid curve). Note that $E$ are the normalized energies.}
\label{dos}
\end{figure}

FEGOE$(k)$[BEGOE$(k)$] are generic although analytically difficult to deal with. The universal properties derived using FEGOE$(k)$[BEGOE$(k)$] extend easily to systems modeled by few-body interactions such as lattice spin models. As the matrix structure of FEGOE$(k)$[BEGOE$(k)$] is different from a classical GOE, there are differences in both one-point (eigenvalue density) and two-point (fluctuation measures) functions. For example, the eigenvalue density is semi-circular for a GOE whereas it is Gaussian for FEGOE$(k)$[BEGOE$(k)$] with sufficiently large particle numbers ($m >> k$) and large $N$. Figure \ref{dos} shows an example. Eigenvalue density for GOE is semicircular whereas FEGOE(2) and BEGOE(2) have approximately Gaussian eigenvalue densities, though they show finite size errors. Here, we make comparison with Edgeworth (ED) corrected Gaussian taking into account corrections due to third (skewness $\gamma_1$) and fourth (kurtosis $\gamma_2$) moments \cite{St-87}, 
\begin{equation}
\rho_{ED}(E) = \displaystyle\frac{1}{\sqrt{2\pi}} \; \exp\left( - \displaystyle\frac{E^2}{2} \right) \left\{  1 + \displaystyle\frac{\gamma_1}{6} He_3(E) + \displaystyle\frac{\gamma_2}{24} He_4(E) + \displaystyle\frac{\gamma_1^2}{72} He_6(E) \right\} \;.
\label{eq-6}
\end{equation} 
Here, $He$ are the Hermite polynomials: $He_3(x) = x^3-3x$, $He_4(x) = x^4-6x^2+3$, and $He_6(x) = x^6-15x^4+45x^2-15$. Also, $E$ are the normalized energies, i.e. centroids are zero and variances are unity. In Figure \ref{dos}, notice deviations at the spectrum edges and in the bulk for FEGOE(2)/BEGOE(2). The eigenvalue density for BEGOE shows slower convergence to Gaussian as $\ell=2$. Thus, the one-point function for FEGOE/BEGOE is different from that of a GOE. 

These results can also be understood in terms of the fourth moment (kurtosis $\gamma_2$) for the eigenvalue density for FEGOE(2) and BEGOE(2). In the dilute limit ($\ell \to \infty$, $m \to \infty$, $m/\ell \to 0$, finite $k$) for fermions, $\gamma_2 \to -4/m$ and $\gamma_2 \to -4/\ell$ for bosons in the dense limit ($\ell \to \infty$, $m \to \infty$, $m/\ell \to \infty$, finite $k$), using binary correlation approximation method \cite{Mo-75}. Thus, BEGOE(2) gives Gaussian eigenvalue densities for sufficiently large values of $\ell$. Similarly, $m > 4$ gives Gaussian eigenvalue densities for FEGOE(2).

By construction, FEGOE($k$)[BEGOE($k$)] has orthogonal invariance in $k$-particle spaces but is not invariant in $m > k$ particle spaces. Moreover, due to the few-body nature of interactions, the many-particle matrix elements of FEGOE($k$)[BEGOE($k$)] are appropriate linear combinations of matrix elements in the defining $k$-particle spaces resulting in non-zero cross-correlations between spectra with different particle numbers $m$ (for fixed $k$) and vice-versa. 
Despite a few attempts \cite{Ve-84,Le-90,Be-01,Be-01a,Sr-02,Be-03,Pa-11}, the two-point correlation function (${\overline{\rho(E_1) \rho(E_2)}} - {\overline{\rho(E_1)}}\;{\overline{\rho(E_2)}}$) for FEGOE($k$)[BEGOE($k$)] could not be derived. For more details, please refer to \cite{Br-81,Gu-98,Be-03,Ko-14}.

\begin{figure}[ht]
  \centerline{\includegraphics[width=6in]{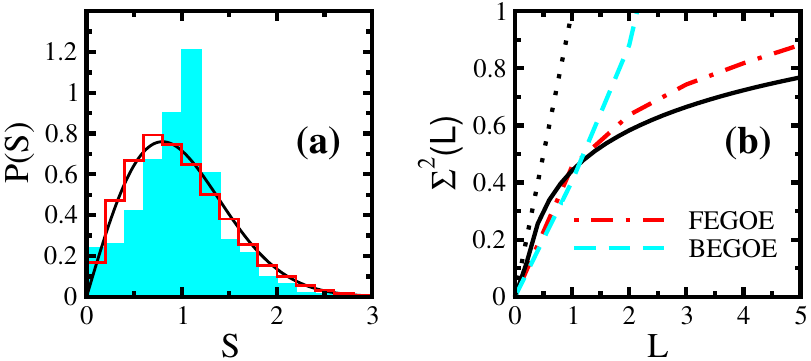}}
  \caption{(a) NNSD for FEGOE and two level BEGOE. The solid histogram for BEGOE and empty histogram for FEGOE are compared with Wigner's surmise for GOE (solid line). (b) Number variance $\Sigma^2$ for FEGOE and two level BEGOE compared with Poisson (dotted line) and GOE result (solid line). The parameters used for constructing FEGOE and two level BEGOE in both panels are same as that in Figure \ref{dos}.}
\label{nnsd-del3}
\end{figure}

For FEGOE$(k)$[BEGOE$(k)$], the fluctuation measures follow GOE using `spectral unfolding' i.e. each member of the ensemble is unfolded individually rather than the 'ensemble unfolding' involving unfolding with ensemble averaged eigenvalue density (defined in Section 2) used for GOE. Figure \ref{nnsd-del3}(a) shows NNSD and Figure \ref{nnsd-del3}(b) shows number variance $\Sigma^2$ for a 100 member FEGOE ($\ell=10$, $m=5$, $k=2$) and BEGOE ($\ell=2$, $m=251$, $k=2$). We have used ED corrected Gaussian (see Eq. \ref{eq-6}) for spectral unfolding. As seen from the NNSD plot, $P(S)$ for FEGOE is pretty close to the Wigner surmise (solid curve). However, for BEGOE, $P(S)$ shows a marked peak close to $S=1$ and $P(S=0) \neq 0$ due to the fact that we consider two level BEGOE although we have $(m >> \ell, m >> k)$. This reflects the situation that we are not too far from a harmonic oscillator (picket fence) statistics \cite{As-01}. Also, $\Sigma^2$ for FEGOE displays logarithmic dependence on $L$ like GOE while for BEGOE, $\Sigma^2$ is almost constant for small $L$ which eventually deviates from GOE and grows with $L$. This may arise due to superposition of sequences of picket-fence like spectra and consequently, some energy levels are almost degenerate \cite{As-01}. With increasing particle numbers (bosons/fermions) and/or increasing number of energy levels, the $\Sigma^2$ for FEGOE and BEGOE shows good convergence to GOE result for small $L$. We have confined ourselves, in next section, to the application of these ensembles (with specific chosen parameters) in studying decoherence of a bipartite quantum system interacting with environment. More recent results by Flores {\it et al} \cite{Fl-01} show that the semi-Poisson distribution $P(S) = 4S\exp(-2S)$ gives a better fit for NNSD in the low-energy part of the spectra generated by two-body interactions, if spectral unfolding is used. Also, the spectral averaged number variance $\Sigma^2_s(L)$ will be different from ensemble averaged number variance $\Sigma^2_e(L)$: $\Sigma^2_s(L) = \left[\Sigma^2_e(L) - (L^2 -\frac{1}{6}) \frac{\sigma^2}{\delta^2} \right] \; \left[ 1-\frac{\sigma^2}{\delta^2}\right]^{-1}$;  $\delta$ denotes the average level spacing and $\sigma^2$ is the ensemble averaged second moment. Here, the correction to $\Sigma^2_e(L)$ is applied after re-centering the spectra to obtain $\Sigma^2_s(L)$. Number variance for two non-interacting particles and $k=1$ has been analyzed recently \cite{Pr-16}.

\begin{figure}[ht]
  \centerline{\includegraphics[width=4in]{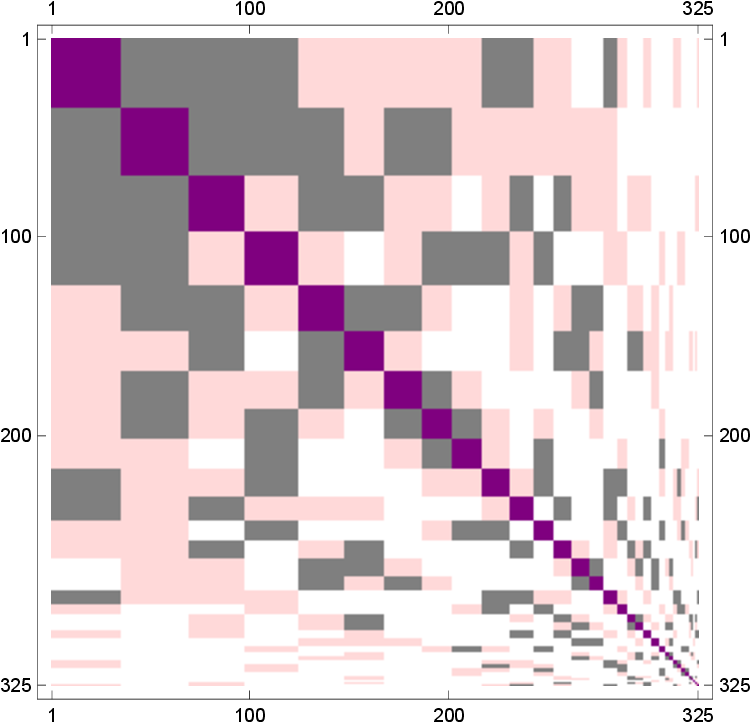}}
  \caption{Hamiltonian structure displaying two-body selection rules. See text for details.}
\label{tbsr}
\end{figure}

It is also of wider interest to understand universality and ergodicity of FEGOE($k$)/BEGOE($k$). A large variety of FEGOE($2$)/BEGOE($2$) with symmetries have been analyzed \cite{Ma-12t} and their analysis show that the FEGOE/BEGOE have universality, i.e. results are generic with wide-ranging applications to physical systems. Ergodicity allows us to compare the ensemble averaged results from random matrix ensembles to spectral averaged results (also with time averages). In the dilute limit for fermions ($\ell \to \infty$, $m \to \infty$, $m/\ell \to 0$, finite $k$) and dense limit for bosons ($\ell \to \infty$, $m \to \infty$, $m/\ell \to \infty$, finite $k$) respectively, FEGOE($k$) and BEGOE($k$) seems to be ergodic. This can be inferred from study of level fluctuations after unfolding. The variance $\Sigma_{11}$ of the centroid fluctuations and the variance $\Sigma_{22}$ of the variance fluctuations are $\Sigma_{11}(m,{m^\prime}) = \overline{\langle H \rangle^m \langle H \rangle^{m^\prime}} - \overline{\langle H \rangle^m} \; \overline{\langle H \rangle^{m^\prime}}$ and $\Sigma_{22}(m,{m^\prime}) = \overline{\langle H^2 \rangle^m \langle H^2 \rangle^{m^\prime}} - \overline{\langle H^2 \rangle^m} \; \overline{\langle H^2 \rangle^{m^\prime}}$. Using binary correlation approximation method, for BEGOE($k$), $\Sigma_{11} \to 4/\ell^2$ and $\Sigma_{22} \to 16/\ell^4$. Therefore, the fluctuations in centroids and variances will tend to zero as $\ell \to \infty$. Moreover, numerical results for fluctuations in $\gamma_1$ and $\gamma_2$ rapidly go to zero with $\ell \to \infty$. Thus, BEGOE($k$) will be ergodic in dense limit defined by ($\ell \to \infty$, $m \to \infty$, $m/\ell \to \infty$, finite $k$). Similarly, for FEGOE($2$), $\Sigma_{11} \to 4mm^\prime/\ell^4$ and $\Sigma_{22} \to 8/\ell^4$ which tends to zero in the dense limit defined by ($\ell \to \infty$, $m \to \infty$, $m/\ell \to 0$, finite $k$). It is important to mention that for two-level BEGOE($k$), the matrix structure is special (tri-diagonal following conventions described above) resulting in non-ergodicity in the dense limit $(m >> \ell, m >> k)$. In general, BEGOE($k$) with finite $\ell$ in the dense limit defined by $(m >> \ell, m >> k)$ will be non-ergodic \cite{As-01}. Also, for unitary case with $\ell=2$ and $m>>k$, the spectrum displays a number of quasi-degenerate states \cite{As-01}. It is important to find experimental signatures for cross-correlations (lower order moments of the two-point function) as they will provide direct evidence for embedded ensembles (they are zero for a classical Gaussian ensemble).

In literature, FEGOE($k$) are also called $k$-BRE \cite{Vo-08} (with acronym BRE for Body Random Ensemble). Simplest of FEGOE($k$)/BEGOE($k$) is the FEGOE($2$)[BEGOE($2$)] generated by random two-body interactions. Note that FEGOE(2) is also called TBRE (Two Body Random Ensemble). Besides two-body interactions, realistic systems also contain an average one-body part and thus, it is more appropriate to use FEGOE(1+2)[BEGOE(1+2)]. It is possible to draw the single particle energies defining the one-body part of the Hamiltonian from the eigenvalues of a random ensemble and then the corresponding FEGOE(1+2) is called two-body random interaction model (TBRIM) \cite{Fl-97} or from the center of a GOE and then the corresponding FEGOE(1+2) is called random interaction matrix model (RIMM) \cite{Al-00,Al-01}. In addition, realistic Hamiltonians carry a variety of symmetries. For example, spin ${\cal S}$ is a good quantum number for mesoscopic sytems, total angular momentum $J$ and parity $\pi$ are good quantum numbers for nuclei, and so on. Therefore, it is important to study FEGOE(1+2)[BEGOE(1+2)] with good symmetries (in addition to the particle number) and this, in principle, provides a systematic classification of embedded ensembles. Many different embedded ensembles with symmetries have been identified and analyzed \cite{Ma-12t} using diversified methods like numerical Monte-Carlo, binary correlation approximation, trace propagation, group theoretical and perturbative methods. 

For FEGOE(2), many of the $m$-particle matrix elements will be zero with only three different types of non-zero matrix elements respectively for zero, one, and two particle transfers. Figure \ref{tbsr} gives an example of block matrix structure of $H$ matrix of $^{24}$Mg displaying two-body selection rules. The total number of blocks is $33$ generating $325$ dimensional $H$ matrix, each labeled by the spherical configurations $(m1,m2,m3)$ obtained by distributing 8 valence nucleons over three spherical orbits $(1d_{5/2},1d_{3/2},2s_{1/2})$. The configurations are ordered such that the blocks are in decreasing order of dimension. The diagonal blocks are shown in purple and within these blocks there will be no change in the occupancy of the nucleons in the three $sd$ orbits. Gray corresponds to the region (in the matrix) connected by the two-body interaction that involve change of occupancy of one nucleon. Similarly, pink corresponds to change of occupancy of two nucleons. Finally, white correspond to the region forbidden by the two-body selection rules. This figure is shown in the spirit of \cite{Ma-12} and a similar figure was given earlier in \cite{Pa-05} for $^{28}$Si with $(J^\pi T ) = (0^+0)$. 

Having understood the one and two point functions, now we proceed to apply the FEGOE and BEGOE to study decoherence of a qubit in the presence of environment.

\section{Decoherence using FEGOE and BEGOE}

\begin{figure}[ht]
  \centerline{\includegraphics[width=6in]{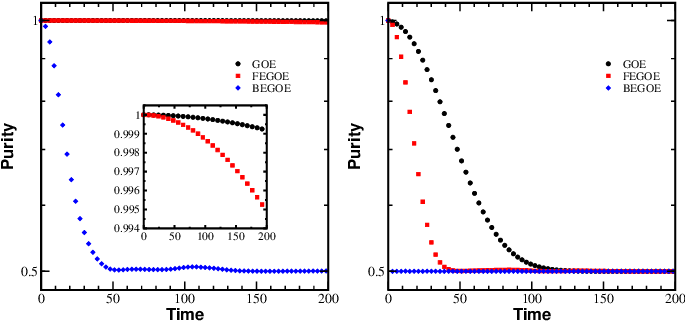}}
  \caption{Ensemble averaged purity of a qubit averaged over a 100 member GOE (circle), FEGOE (square) and BEGOE (diamond) environments with coupling strength $\lambda=10^{-4}$ (left) and $\lambda=10^{-2}$ (right). The dimension of environment in all the cases is $252$. Inset in the left picture shows the zoomed-up part for GOE and FEGOE.}
\label{purity}
\end{figure}

The field of applications of embedded ensembles has grown in recent years in parallel with the growth of applications of RMT in many-body physics. We shall concentrate here on one aspect very relevant to this year's ELAF. This is the application to decoherence \cite{Go-06, Pi-07, Mar-07, Go-08, Da-09, Mar-09, Le-11, Pi-15, Mo-15, Gr-17, Br-17}. In all these papers the environment, or one of them, is represented by a classical random matrix ensemble. Yet in all these cases it actually is a many-body system. Note that an alternative was proposed in \cite{Go-16}. The environment, as a many-body system can be represented in a tight binding aproximation, which in turn can be mimicked by a random spin network or some other quantum graph giving rise to a DEGOE. Analytically little is known except that complete graphs yield the universal results of classical ensembles for the joint probablility distribution \cite{Pl-15}. Yet such models allow rather simple numerical calculations. We shall not consider them in the present paper.

We here will rather assume that we have a realistic system of many fermions or bosons and describe them by FEGOE and BEGOE. We will emphasize both differences and similarities of the two cases. To achieve this, we shall study the unitary dynamics of a bipartite quantum system namely one qubit (q) under the influence of an environment (e). The environment is finite-dimensional, random and is modeled by GOE, FEGOE and BEGOE. The Hilbert space structure is given by ${\cal H} = {\cal H}_q \otimes {\cal H}_e$ and the dynamics is governed by the Hamiltonian,
\begin{equation}
H = H_q \otimes {\mathds{1}}_e + {\mathds{1}}_q \otimes H_e + \lambda \; \sigma_z \otimes V_e \;.
\label{ham-dec}
\end{equation}
In Equation \ref{ham-dec}, $H_q=\sigma_z/2$ (the level spacing is unity) and the positive parameter $\lambda$ is the coupling strength between the qubit and the environment. Notice that the coupling term (third term in Equation (\ref{ham-dec})) is chosen to be separable, with each operator acting on the corresponding Hilbert space. The matrix $\sigma_z$ is one of the Pauli matrices $(\sigma_x, \sigma_y, \sigma_z)$. As $\sigma_z$ commutes with $H_q$, we have chosen dephasing coupling \cite{Ga-97}. The average level spacing in the spectrum of $H_e$ is chosen to be unity. With $\hbar=1$, time is measured in terms of Heisenberg time $t_{\mathrm{H}} = 2\pi$ and the time evolution is given by the operator $U(t) = \exp(-iHt)$. The initial density matrix for the total system is $\rho(0) = \rho_q(0) \otimes \rho_e(0)$. At any given time $t$, the reduced dynamics for qubit is obtained by tracing out the environment, $\rho_q(t) = \mathrm{Tr}_e[\exp(-iHt) \; \rho(0) \; \exp(iHt)]$. Decoherence is quantified in terms of purity $P(t) = \mathrm{Tr}[ \overline{\rho_q^2(t)}]$.

Our model contains two random matrices, the environment Hamiltonian $H_e$ and the environment operator in the coupling term $V_e$. The matrix $H_e$ is a diagonal matrix of eigenvalues of a random matrix unfolded to unit average level spacing across full spectrum length. The initial state is a separable pure state, $\Psi=\Psi_q \otimes \Psi_e$. For the qubit, the initial state is chosen to be a symmetric eigenstate of $\sigma_x$, $\Psi_q = (| 0 \rangle +  | 1 \rangle)/\sqrt{2}$. Trivially, there will be no dynamics if initial state is chosen to be an eigenstate of $\sigma_z$. The environment part $\Psi_e$ is given by a random state which is invariant under orthogonal transformations. We make following choices for $H_e$ and $V_e$: (a) $H_e \in$ GOE and $V_e \in$ GOE, (b) $H_e \in$ FEGOE(2) and $V_e \in$ FEGOE(1), and (c) $H_e \in$ BEGOE(2) and $V_e \in$ BEGOE(1). 

Figure \ref{purity} shows numerical results for the purity $P(t)$ of the qubit for the three choices of $H_e$ and $V_e$ mentioned above. We make following choice of parameters for FEGOE: $\ell=10$, $m=5$ and BEGOE: $\ell=2$, $m=251$. The dimension of environment for all the cases is $252$ and we generate a 100 member ensemble in each case. For smaller coupling strength $\lambda$, the purity decay is slowest for GOE environment. It is comparatively faster for FEGOE environment and fastest for two level BEGOE environment.
The dynamics is faster when we increase the coupling strength from $0.0001$ (left panel) to $0.01$ (right panel). For $\lambda=0.01$, the purity for two level BEGOE almost instantaneously saturates which can be attributed to its Hamiltonian structure and moreover, two-level bosonic systems are integrable irrespective of $k$ in the semi-clasical limit \cite{Be-03a}. Our calculations bring out the dependence of decoherence on the nature of the environment very clearly. The rank of the interactions affects the rate of decay of purity and this can be understood in terms of spectral variances but this is for future. Detailed analysis of decoherence of a qubit system in presence of random FEGOE/BEGOE environments will be presented in a separate paper.

\section{CONCLUSIONS AND OUTLOOK}

We hope to have given a brief overview of the critical problems of embedded ensembles,
and also of their fundamental importance. First of all, we have the problem of ergodicity, which is still open, but which is known to be of marginal relevance as convergence in any case would be extremely slow. On the other hand, we have shown that a true necessity for embedded ensembles exists, because in relevant cases deviation from the classical ensembles can be very large. Also we have shown that numerical methods can give rather satisfactory results, because often our problems are of finite dimension and this can be handled by mixing numerical studies with known physics and analytical approximations.

\section{ACKNOWLEDGMENTS}
Authors acknowledge supercomputing facility LANCAD-UNAM-DGTIC-330 and financial support from UNAM/DGAPA/PAPIIT research grant IA104617 and CONACyT research grant 219993.


\bibliographystyle{aipnum-cp}%
\bibliography{pap-ELAF17-revised}%

\end{document}